\documentclass[aps,nofootinbib]{revtex4}
\usepackage{epsfig}
\usepackage{float,amsmath,amstext,amsthm,amssymb,amsfonts}
\usepackage{graphicx}

\begin{document}

\title{Thermalization of gluons and onset of collectivity
at RHIC due to $gg \leftrightarrow ggg $ interactions}

\author{C.~Greiner$^{1}$ and Z.~Xu$^{1}$}

\affiliation{$^{1}$Institut f\"ur Theoretische Physik, %
   Johann Wolfgang Goethe Universit\"at Frankfurt, %
   Max-von-Laue-Str.~1, %
   D--60438 Frankfurt, %
   Germany \vspace*{5mm}}

\begin{abstract}
A number of promising results of a new 3+1 dimensional 
Monte Carlo parton cascade including 
inelastic multiplication
processes ($gg\leftrightarrow ggg$) are elaborated:
(1) thermalization and chemical saturation;
(2) the onset of longitudinal hydrodynamical expansion;
(3) the build up of elliptic flow. 
We first briefly outline the basic idea of the algorithm. 
Full simulations are done with initial conditions for
the kinetic partons via minijets
or with ones stemming from a color glass condensate.  
The inclusion of the inelastic channels
leads to a very fast kinetic equilibration and also
to an early creation of pressure.

\end{abstract}

\maketitle


\section{Introduction and brief description of the cascade }

\label{Intro}
The prime intention for present ultrarelativistic heavy ion collisions
at CERN and at Brookhaven lies in the possible experimental identification of
a new state of matter, 
the quark gluon plasma (QGP). Measurements \cite{Exp} at RHIC
of the elliptic
flow parameter $v_2$ for nearly central collisions suggest that - in comparison
to fits based on simple ideal hydrodynamical models - the
evolving system builds up a sufficiently  early
pressure and potentially also achieves (local) equilibrium.

On the other hand,   
the system in the reaction is at least initially
far from any equilibrium configuration.
To microscopically describe and understand the dynamics of 
ultrarelativistic heavy ion collisions,
and to address the crucial question of thermalization and early
pressure buildup, we have developed a kinetic parton cascade algorithm
\cite{Xu04}
inspired by perturbative QCD including
for the first time inelastic (`Bremsstrahlung')
collisions $gg \leftrightarrow ggg $
besides the binary elastic
collisions. 

It is well known, that a parton cascade analysis, incorporating only
elastic (and forward directed) $2 \leftrightarrow 2$
collisions described via one-gluon exchange,
shows that thermalization and early (quasi-)hydrodynamical behaviour
(for achieving sufficient elliptic flow)
can not be built up or maintained,
but only if a much higher, constant and isotropic cross section 
$\sigma _{eff} \approx 45$ mb
is being employed \cite{MG02}.
By employing such a high cross section, however, 
especially for the very early phase when the system
is very dense, the physical justification
of a quasiclassical kinetic transport equation becomes
unwarranted: 
The mean free path $\lambda_{mfp} \approx 1/(n \sigma_{eff})
\sim 1/n $, whereas the mean distance among the gluons is
$d = 1/(n)^{1/3}$. $n$ denotes the number density of gluons.
If $\sigma_{eff} $ stays constant at such a large value and if
the
gluon density is getting large, as typically achieved 
in the early stages of the reaction, one has the unphysical picture that
the mean free path would be much smaller than the interparticle
distance  $\lambda_{mfp} \ll d $. The use of a semiclassical, 
kinetic Boltzmann transport
description is, from a classical point of view, unjustified and not valid. 
In quantum mechanical terms, the gluons as dynamical
degrees of freedom would aquire a collional width 
$\Gamma \approx n \sigma_{eff} v_{rel} \approx 1/\lambda_{mfp} $ being then
much larger than the typical energy $E \approx 3 T \sim 1/d$,
and thus would resemble very broad excitations and are not 
quasi-particles by any means. 
On the other hand, typical Debye screened
pQCD cross section scale roughly like the inverse temperature squared,
$\sim 1/T^2$, so that $\lambda_{mfp} $ does not become smaller than $d$,
but is of the same order and slightly larger. This is also
true for inelastic channels, as long as the coupling stays small enough.

In addition, 
the possible importance of the inelastic reactions
on overall thermalization was
raised in the so called `bottom up thermalization' picture \cite{B01}.
It is intuitively clear that gluon multiplication should
not only lead to chemical equilibration, but
also should lead to a faster kinetic equilibration.
This represents a further (but not all) important motivation
for developing a consistent algorithm to handle
also inelastic processes.

The conceptual new simulation lies in 
treating elastic and inelastic multiplication collisions
in a unified manner \cite{Xu04}.
Most importantly, the (multiparticle) back reation channel
($ggg\rightarrow gg$) is treated fully consistently
by respecting detailed balance within the same 
algorithm. If the back channel would be neglected,
one would possess no serious handle on soft gluon production!
The system would cool too fast simply by gluon multiplication,
and the gluon population (`entropy' production) would dramatically
oversaturate. This states a serious problem in the
older parton cascade schemes \cite{VNI}, where standard
gluon splitting without adequate recombination is employed.

The numerical challenge is how to 
numerically incorporate the fusion \cite{Xu04}:
The idea of the stochatic method is to divide the total space in 
sufficiently small local cells,
in which the system is quasi homogenous. In this local cells
master equations in momentum space are solved.
Detailing on the the back reaction of $gg \leftrightarrow ggg$, we define
a transition probability in a time interval
with $0\leq P_{32} \ll 1$ for a given triplet of gluons
with specific momenta 
in a small local cell:
\begin{equation}
\label{p32}
P_{32} = \frac{\Delta N_{coll}^{3\to 2}}{\Delta N_1 \Delta N_2 \Delta N_3}
= \frac{1}{8E_1 E_2 E_3} \frac{I_{32}}{N_{test}^2}
\frac{\Delta t}{(\Delta^3 x)^2} \, ,
\end{equation}
where $I_{32}$ is defined
as the phase space integral 
$$
\frac{1}{2!} \int \frac{d^3 p^{'}_1}{(2\pi)^3 2E^{'}_1}
\frac{d^3 p^{'}_2}{(2\pi)^3 2E^{'}_2} | {\cal M}_{123\to 1'2'} |^2 (2\pi)^4
\delta^{(4)} (p_1+p_2+p_3-p^{'}_1-p^{'}_2) \, .
$$  
$\Delta^3 x $ denotes the volume of a specific cell and
$N_{test}$ represents the number of test-particles.
The entering matrix element one obtains
from the $2\rightarrow 3$ element via a standard
prefactor governed by a detailed balance relation.
The latter is given by
\begin{equation}
\label{m23}
 | {\cal M}_{gg \to ggg} |^2 = ( \frac{9 g^4}{2} 
\frac{s^2}{({\bf q}_{\perp}^2+m_D^2)^2}  ) 
( \frac{12 g^2 {\bf q}_{\perp}^2}
{{\bf k}_{\perp}^2 [({\bf k}_{\perp}-{\bf q}_{\perp})^2+m_D^2]}  ) 
\theta (k_{\perp} \Lambda_g - \cosh y ) \,  ,
\end{equation}
where $g^2=4\pi\alpha_s$.
${\bf q}_{\perp}$ and ${\bf k}_{\perp}$ are the perpendicular component of
the momentum transfer and that of the momentum of the radiated gluon 
in the c.m.-frame of the collision, respectively.
$y$ denotes the rapidity of the
radiated gluon.
We thus take 
$gg \rightarrow ggg $
in leading-order of pQCD and
consider an effective Landau-Pomeranchuk-Migdal suppression
with $\Lambda_g$ denoting the gluon mean free
path, which is given by the inverse of the total gluon 
collision rate $\Lambda_g=1/R_g$,
and also employ a standard screening mass $m_D$
for the infrared sector of the
scattering amplitude. Both $m_D$ and $\Lambda_g$ are calculated
selfconsistently. 
The coupling is taken as scale dependent.
If the system is undersaturated, accordingly, the
cross sections than are noticeably higher than compared to the magnitude at 
thermal equilibrium.

Incorporating the algorithm in a full
3+1 dimensional Monte Carlo cascade, one  
achieves a covariant parton cascade which can accurately handle
the immense elastic as well as inelastic scattering rates 
occuring inside the dense (gluonic) system.
The important task is to develop a dynamical mesh of
(expanding) cells in order to handle the extreme
initial situation.
For an exhaustive testing of the code we refer to the original
paper \cite{Xu04}. Instead, in the following section,
we give a number of, as we believe, important and to 
some extent also still rather preliminary results
obtained via real 3+1-dim. simulations for RHIC energies.

\section{Selected Results of the cascade operating at RHIC}
\label{PCtherm}

We first address
the question of the importantance of the (still) pQCD inspired reactions
on the thermalization and early
pressure build up for heavy ion collisions at RHIC.
The algorithm can incorporate
any specified initial conditions
for the freed on-shell partons.
The first results we show  
take as a conservative point of view 
minijet initial conditions.

\begin{figure}[ht]
\vspace*{0.1cm}
\includegraphics[width = 3.5in ]{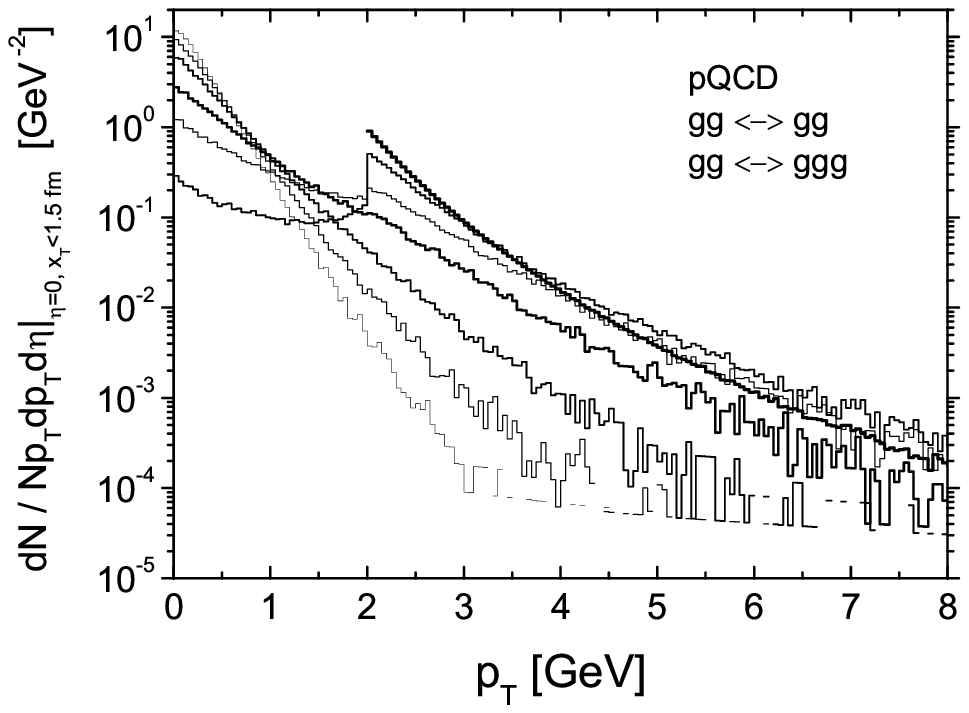}
\includegraphics[width = 3.5in ]{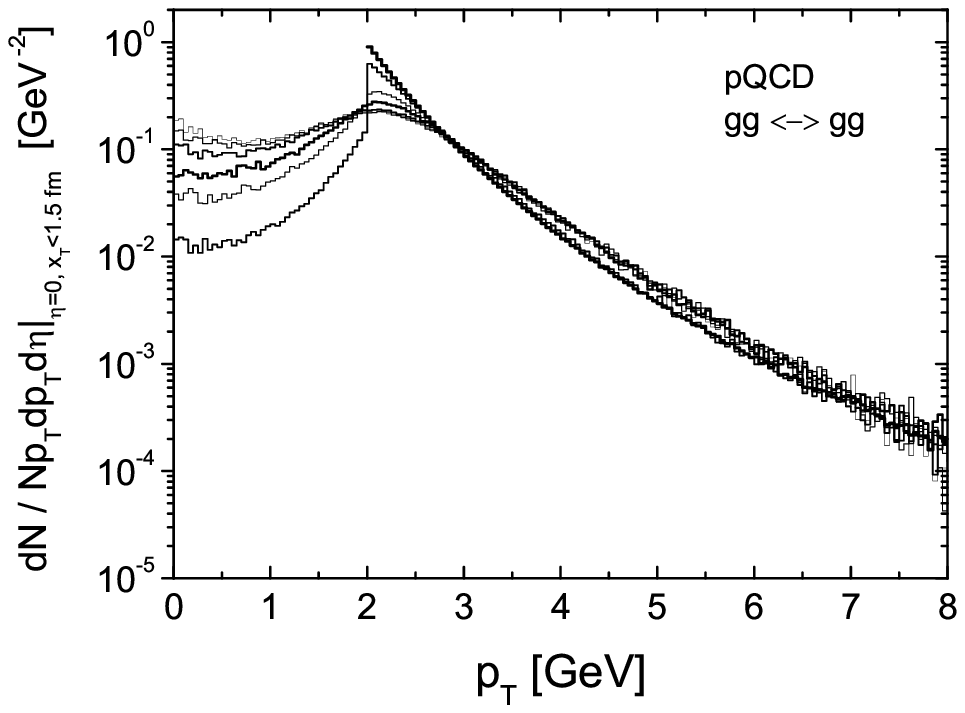}
\vspace*{-0.6cm}
\caption[]{Left panel:
Transverse momentum spectrum at
(spatial) midrapidity ($\Delta \eta =1$) at different times
(t=0.2, 0.5, 1, 2, 3 and 4 fm/c
from second upper to lowest line) 
for a real, central
fully 3-D 
ultrarelativistic heavy ion collision. Only the partons residing
in a central cylinder of radius $R\leq 1.5$ fm are depicted.
From $t=0$ on first only gluonic minijets with $p_t>2$ GeV are populated
(most-upper, boldfolded line).
Energy degradation to lower momenta proceeds rapidly by
gluon emission within the first fm/c. Maintenance of
(quasi-)kinetic and (later) chemical equilibrium is given up to 4 fm/c,
where longitudinal and transversal
(quasi-hydrodynamical) work is done resulting in a continous
lowering of the temperature.
 Right panel: Like the left panel, but now 
by incorporating only elastic pQCD collisions.
The most-upper and boldfolded histogram with a lower cutoff
at $p_T=2$ GeV denotes the spectrum of the primary gluons
(minijets).
}\label{dn}
\end{figure}

Minijet production comes from multiple binary 
nucleon-nucleon-scattering in a nucleus-nucleus-collision,
where we have chosen a conservatively large
transverse momentum cutoff of $p_t > p_{0} = 2$ GeV/c 
\cite{Es89},
according to the differential jet cross section:
\begin{equation}
\label{csjet}
\frac{d\sigma_{jet}}{dp_T^2dy_1dy_2} = K \sum_{a,b}
x_1f_a(x_1,p_T^2)x_2f_b(x_2,p_T^2) \frac{d\sigma_{ab}}{d\hat t} \, .
\end{equation} 
$p_T$ denotes the transverse momentum and $y_1$ and $y_2$ are the rapidities
of the produced partons.
The partons are 
are distributed in space-time via the corresponding overlap function.
In the original paper we have chosen no formation time for
the minijet gluons, so that they can immediately interact. One can think of
a phenomenological formation time 
$\tau_f \approx 1/p_t $ for the gluons to become
onshell and freed from the nucleus wavefunction. 
For the first two figures we show
(taken from \cite{Xu04})
no formation time is employed. The ones after are calculated 
by incorporating such a formation time.
As it turns out, when doing calculation w/w.o. formation time the difference 
in extracting quantities is really small, after the very early formation 
of minijet gluons has terminated.

\begin{figure}[htb]
\begin{center}
        \includegraphics[width=9cm]{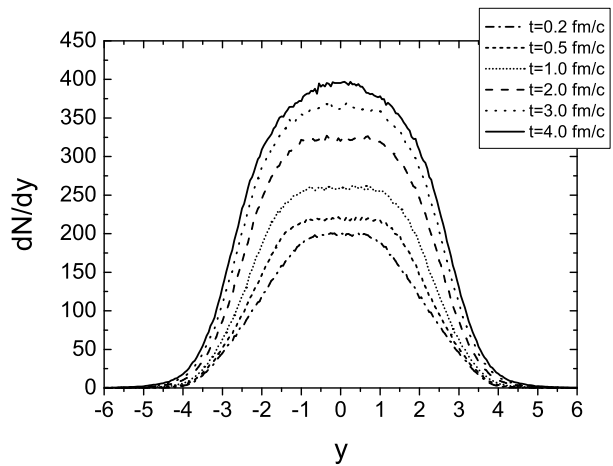}
\caption[]{Gluon number distribution versus momentum rapidity at
the time $t=0.2$, $0.5$, $1.0$, $2.0$, $3.0$ and $4.0$ fm/c during the
expansion.
}
\label{dndy}
\end{center}
\end{figure}

For a cutoff of  $p_{0} = 2$ GeV/c the number of initial minijets
is about 900 with a $\frac{dN_g}{dy}\approx 200$ at midrapidity
for central Au+Au collisions at $\sqrt{s}=200 $ AGeV. 
This is a rather low value, but we keep to this conservative 
estimate for our first two figures. 
In Figs. \ref{dn} we show the 
transverse momentum spectrum obtained with the full dynamics at
(spatial) midrapidity ($\Delta \eta =1$) at different times
for partons of a central cylinder of radius $R\leq 1.5$ 
and, respectively, a similar calculation with including
only elastic collisions.
From $t=0$ on first only gluonic minijets with $p_t>2$ GeV are populated.
Energy degradation to lower momenta proceeds rapidly by
gluon emission within the first fm/c. Maintenance of
(quasi-)kinetic and (later) chemical equilibrium is given up to 4 fm/c,
where longitudinal and transversal
(quasi-hydrodynamical) work is done resulting in a continous
lowering of the temperature.
This can be seen by the continous steepening of the exponential
slopes of the spectrum with progressing times.
It turns out that kinetic momentum equilibration occurs at times
of about 1 fm/c, whereas full chemical equilibration occurs
on a smaller scale of about 2-3 fm/c.
As the right panel of the figure does not show any sign
of strong momentum degradation, thermalization 
is clearly due to the
incoporation of the inelastic channels.  
For the complete transversal region
a remedy of the initial non-equilibrium high momentum tail will remain
stemming from the escaping minijets of the outer region.

\begin{figure}[htb]
\begin{center}
        \includegraphics[width=12cm]{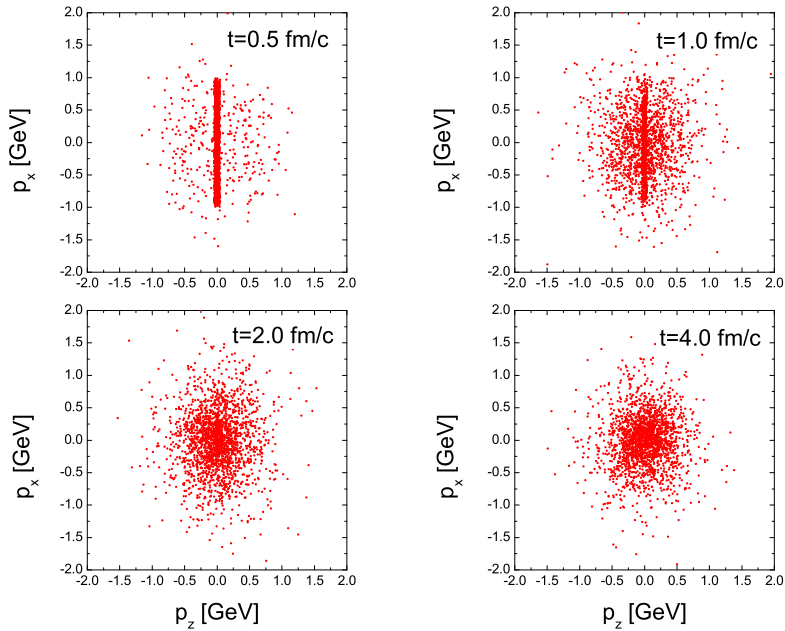}
\caption[]{`bottom up thermalization': A typical color glass
condensate initial condition is taken at $\tau_0=0.4$ fm/c
within a Bjorken geometry. The saturation scale is taken as
$Q_s=1$ GeV. The evolution of the momentum occupation
is shown for four subsequent times sampled 
within a space-time rapidity
interval of $\Delta \eta = 0.1$ and a central transverse region 
of $R\leq 1.5$ fm. One clearly recognizes the population 
of the `soft' gluons and a subsequent degradation of
the `hard' initial gluons.
}
\label{cgc}
\end{center}
\end{figure}

An intuitive interpretation of the fast thermalization is the following:
Whereas the elastic scattering is forward peaked, this
is not fully the case for the inelastic reaction including a LPM cutoff
\cite{Xu04}.
Especially the emitted gluons show a flat and non-forward 
angle distribution. This underlines why the inelastic processes
are so important not only for accounting for chemical equilibration,
but also for kinetic equilibration. It are these Bremsstrahlung radiations 
which actually bring about early thermalization to the QGP.

In Fig. \ref{dndy} the ongoing production of gluons versus rapidity is
given. The reason is that the system, 
for the momentum cutoff chosen, starts being highly undersaturated
in gluon number so that abundant gluon production by Bremsstrahlung
sets in right after the minijets have appeared. 
At the end of the evolution at t=4 fm/c about
$\frac{dN_g}{dy} \approx 400$ are at midrapidity, i.e. the gluon number has
doubled. If we compare the amount of transversal energy
at midrapidity with the experimental factor, we are roughly below
by at least a factor of 2. This means, that for the initial conditions
too few gluons have been assumed.

In Fig. \ref{cgc} we now give a first dynamical realization
of the so called `bottom up thermalization' scenario 
as advocated in \cite{B01}. The initial distribution of gluons
is taken as that of a characteristic color glass condensate
with the same parameters as in 
the simple parametrization given in \cite{BjoRa}. The initial geometry
is of Bjorken type. The evolution of the momentum occupation
is shown for four subsequent times sampled 
within a space-time rapidity
interval of $\Delta \eta = 0.1$ and a central transverse region 
of $R\leq 1.5$ fm. One nicely recognizes the population 
of the `soft' gluons at an early time scale and a subsequent degradation of
the `hard' initial gluons. All this happens roughly within 1 fm/c.
In the last picture we also see that all particles are clearly
more centered around the origin, demonstrating once more
the ongoing cooling and quasi hydrodynamical behaviour from 2 to 4 fm/c.
As a surprise, there is one striking difference 
compared to the idealistic scenario
of \cite{B01}: The number of gluons (per unit rapidity) is slighly
decreasing, although a strong parametric enhancement has been advocated
in \cite{B01} due to Bremsstrahlung production. The reason is that
for the initial conditions taken from \cite{BjoRa}, a clear 
separation of hard and soft scale is not really given, 
the parametric estimate has thus taken to be with caution.
In any case our exploratory study 
of the `bottom up thermalization' picture is interesting 
in its own right and deserves 
further detailed analysis.

\begin{figure}[htb]
\begin{center}
        \includegraphics[width=10cm]{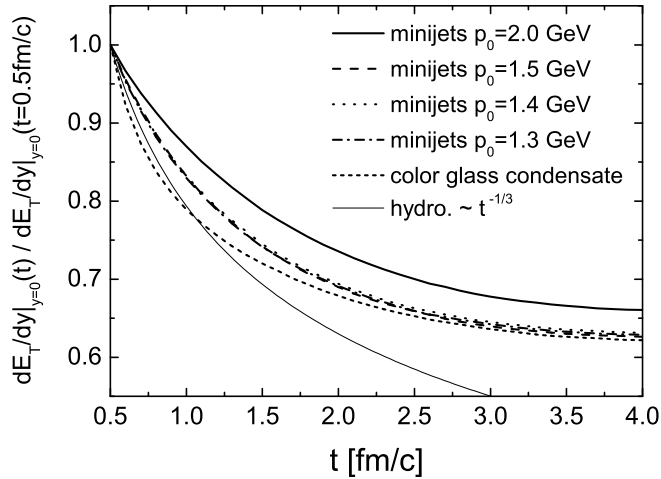}
\caption[]{Time evolution of the transverse energy per unit
momentum rapidity at midrapidity for various initial minijet
conditions and one color glass condensate condition
of central Au+Au collisions at $\sqrt{s}=200 $ AGeV.
All transverse energies are gauged according to their value
at $t=0.5$ fm/c. As a comparison the simple scaling 
of a longitudinally expanding ideal fluid is also given.
For the first 1 to 1.5 fm/c the system clearly expands
close to ideal by exhibiting longitudinal work.  
}
\label{ety}
\end{center}
\end{figure}

Fig. \ref{ety} summarizes our finding with respect to
a potential onset of a collective, longitudinal expansion \cite{pub1}.
Various initial conditions are investigated. For the initial minijet scenario
the cutoff scale $p_0 $ has been lowered from 2.0 down to 1.3 GeV.
(In all this calculation the minijet gluons are now formed with an initial
formation time $\tau_f = 1/p_t$.) The color glass condensate initial
condition is taken as just described above.
In the figure the time evolution of the transverse energy per unit
momentum rapidity at midrapidity is depicted.
All transverse energies are gauged according to their value
at $t=0.5$ fm/c. As a comparison the simple scaling 
of a longitudinally expanding ideal fluid is also given.
For the first 1 to 1.5 fm/c the system expands
close to ideal by exhibiting longitudinal work 
with a ongoing cooling. 
This nicely demonstrates that the system experiences 
quasi ideal hydrodynamic behaviour 
from $t\approx 0.5 $ fm/c to $ t\approx 2$ fm/c, after the gluons
have started to kinetically equilibrate.
If there would be no inelastic pQCD collisions, but only elastic,
the transverse energy would stay almost constant (\cite{Xu04}).
In addition, the evolution of the transverse energy 
seems to be only moderate sensitive to the initial condition
chosen (minijets or color glass condensate), 
showing quite an impressive  universal behaviour \cite{pub1}.  
This behaviour is due to the presence 
of inelastic collisions.

One can also now seek for the `optimal' initial
conditions being in line with RHIC data: From STAR one has  
$\left. dE_T/dy \right| _{y=0} \approx 620 $ GeV for
central collisions.
With the conservative parameter $p_0 = 2$ GeV for the minijet distribution
the final transverse energy would be smaller by a factor of 2.
A similar `underestimation' 
holds for the chosen color glass condensate initial condition, which
thus will need some more special fine-tuning 
(potentially with some high momentum tail, see eg \cite{Hirano} ) 
in order to fit
to data.
For the minijets the `optimal' parametrisation would be choosing
$p_0 \approx 1.4 $ GeV. In such a situation the gluon number
is always close to full saturation.

\begin{figure}[htb]
\begin{center}
        \includegraphics[width=10cm]{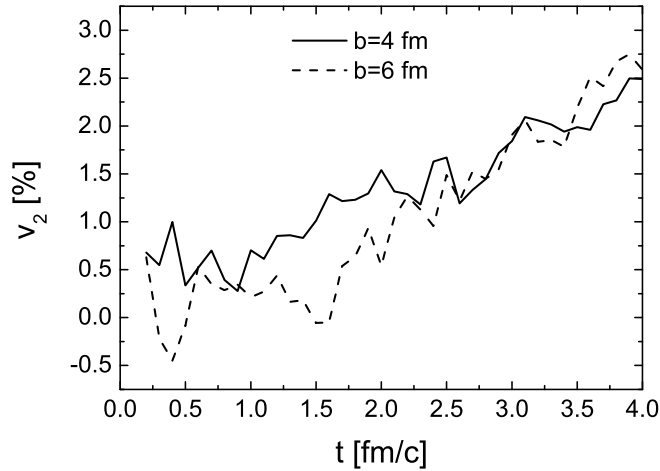}
\caption[]{First and preliminary results for extracting
the elliptic flow parameter $v_2(t)$ at midrapidity
($\Delta \eta =1$) for two noncentral Au+Au collisions 
at an impact parameter b=4 fm and b=6 fm.
As initial conditions minijets have been taken with a 
cutoff $p_0 = 1.5$ GeV. One clearly sees that the flow
steadily builds up to approximately 3 percent 
for the two impact parameters, but
is still not saturated at a time of 4 fm/c.
}
\label{v2prelim}
\end{center}
\end{figure}

In Fig. \ref{v2prelim} we now turn to an important benchmark,
the calculation of the elliptic flow parameter $v_2$.
These calculations are still very preliminary and no exhaustive
testing has been finished. 
As initial conditions minijets have been taken with a 
cutoff $p_0 = 1.5$ GeV. 
We depict the evolution of
$v_2(t)$ of all gluons at midrapidity
($\Delta y =1$) for two noncentral collisions 
at an impact parameter b=4 fm and b=6 fm.
Typically  $v_2$ has its maximum
for an impact parameter of $b=8-10 $ fm with $v_2 \approx 0.06$;
for $b=4$ fm one finds $v_2 \approx 0.03 $ and for $b=6 $ fm
$v_2 \approx 0.04 $
(see eg calculations
and nice compilation and comparison to STAR data in ref. \cite{Lin02}).
The reason why we choose the smaller impact parameter is that
we simply wanted to stay to the default transverse grid of 
subcells. For larger impact parameter a more fine tuned grid
is probably needed and accordingly has to be numerically tested. 
We stop the evolution at times $t=4 $ fm/c as then the
gluon density drops roughly below $1$ fm$^{-3}$, i.e. 
hadronization should set in.
One clearly sees that the flow
steadily builds up to approximately 3 percent 
for the two impact parameters, i.e. $v_2 \approx 0.03 $, but
is still not saturated at a time of 4 fm/c.
These values are already quite large and do come very close to 
the just cited `experimental' results. For
the $b=6 $ fm case the flow might be too low by 20 percent, yet
$v_2(t)$ is still in the tendency of rising with progressing late times.
In addition, a smaller amount of $v_2$ can come from the later
hadronic phase \cite{Bleicher} and/or from the very early
stage of a color glass condensate \cite{Krasnitz}.
It might also be conceivable that preisotropization of the particles
\cite{Dumitru} due to temporary instabilities 
being manifested by strong classical chromoelectromagnetic fields
also will yield some small initial elliptic flow. 
There is still a lot room for investigations,
yet our preliminary results are very encouraging.


\section{Summary and Conclusions}
\label{conclusion}
  
The presented extensive study of
a new and complex parton cascade shows  
that gluon multiplication
via Bremsstrahlung (and absorption) is of utmost importance
to understand kinetic equilibration, chemical saturation
and the build up of early pressure.
The existence of the latter nicely shows up in the continous
steepening of the transverse momentum spectra and the build up
of longitudinal work.
For the various settings of initial conditions
kinetic equilibration is achievd on a timescale of less than about 1 fm/c,
whereas the full chemical equilibration occurs on a somewhat
slower scale of about 2-3 fm/c, if the initial conditions are chosen so
that the gluon number is initially undersaturated.
First results have been presented with color glass condensate initial 
condition. The bottom up thermalization picture seems to work
for a realistic coupling with a kinetic equilibration occuring
again also in less than 1 fm/c. On the other hand no strong amplification
in the gluon number occurs. Finally very preliminary 
results on the build up of elliptic flow have been shown.
Maybe too early to fully claim, it seems   
that approximately 80 per cent of the total
$v_2$ can be induced by the inelastic parton interactions.

Is the QGP a strongly coupled system, a sQGP, 
as advertised in various recent
agenda, where eg high cross sections have been discussed and employed, 
in order to come close to ideal hydrodynamics?
Our analysis is still ongoing and has to be more detailed, 
before claiming it can account for a variety of data.
The first calculation are indeed very encouraging to proceed.
The pQCD cross sections employed are typically on the order
of less than 1 mb to a few mb (at the later stage of the reaction, 
when the system has to hadronize) \cite{Xu04}, depending
also on the degree of gluon saturation. 

In the future a lot of further details have to be
explored \cite{pub2}: Thermalization, also of the
light and heavy quark degrees of freedom, has to be
investigated for
various initial conditions (minijets, Pythia events,
color glass condensate)  
with a detailed comparison to data.
Furthermore we will investigate the full impact parameter dependence
of the transverse energy in order to understand elliptic and transverse flow
at RHIC. Can the inelastic interactions generate almost 
the seen elliptic flow $v_2$, as implied by the exemplaric first
calculations? 
How close are the calculations compared to (ideal) hydrodynamics?
How close to reality? Also the partonic jet-quenching picture can be analysed
in 3-D details. One can also compare the present
calculations with some fixed and specified hydrodynamical initial
conditions directly with calculations based on viscous
relativistic hydrodynamics, either assuming
Bjorken boost invariance within an expanding tube or for full
3+1 dimensions. Such a comparison can tell how viscous
the QGP really turns out to be.

The ultimate aim is to obtain
a consistent picture of the (kinetic) QGP dynamics and to
potentially `deduce' the optimal initial condition
of freed partons. If succesful, one can than extrapolate
to future experiments at the much higher energies at the LHC.

\vfill\eject
\end{document}